\documentclass[aps,prb,twocolumn,showpacs,amsmath,amssymb,superscriptaddress]{revtex4-2}
\usepackage{graphicx}
\usepackage{CJK}
\usepackage{color}
\usepackage[colorlinks,bookmarks=false,citecolor=blue,linkcolor=red,urlcolor=blue]{hyperref}
\usepackage{multirow}
\usepackage{ulem}
\usepackage{tabularx}
\usepackage[nounderscore]{syntax}
\graphicspath{{figs/}}

\newcommand{\be}{\begin{equation}}
\newcommand{\ee}{\end{equation}}

\begin{document}
%\begin{CJK*}{UTF8}{gbsn}
\title{Bulk and Surface Critical Behaviors of quantum Heisenberg antiferromagnet on a two-dimensional coupled diagonal ladders}

\author{Zhe Wang}
\affiliation{Department of Physics, Beijing Normal University, Beijing 100875, China}

\author{Fan Zhang}
\affiliation{Department of Physics, Beijing Normal University, Beijing 100875, China}

\author{Wenan Guo}
\email{waguo@bnu.edu.cn}
\affiliation{Department of Physics, Beijing Normal University, Beijing 100875, China}
\affiliation{Beijing Computational Science Research Center, Beijing 100193, China}

\date{\today}
\begin{abstract}
 Using Quantum Monte Carlo simulations, we study the spin-1/2 Heisenberg model on a two-dimensional lattice formed by 
 coupling diagonal ladders. The model hosts an antiferromagnetic N\'eel phase, a rung singlet product phase, and a
 topological none trivial Haldane phase, separated by two quantum phase transitions. We show that the two quantum 
 critical points are all in the three-dimensional O(3) universality class. The properties of the two gapped
 phases, including the finite-size behavior of the string orders in the Haldane phase, are studied. We show that the surface formed 
 by the ladders ends is gapless, while the surface exposed along the ladders is gapful, in the Haldane phase.
 Conversely, in the gapped rung singlet phase, the former surface is gapped, and the latter is gapless. We demonstrate
 that, although mechanisms of the two gapless modes are different, nonordinary surface critical behaviors are realized 
 at both critical points on the gapless surfaces exposed by simply cutting bonds without fine-tuning the surface 
 coupling required to reach a multicritical point in classical models. We also show that, on the gapped surfaces, the surface critical behaviors are in the ordinary class.
\end{abstract}
\maketitle

%%%%%%%%%%%%%%%%%%%%%%%%%%%%%%%%%%%%%%%%%%%%%%%%%%%%%%%%%%%%

	\section{Introduction}
	\label{intro}

	At a bulk critical point, the surface may show rich and novel critical behaviors, called the surface critical behavior (SCB).\cite{Cardy}
The surface critical behavior is classified into three universality classes: the ordinary, the extraordinary, and the special. 
Typically, the surface orders simultaneously with the bulk, and
the surface singularities are purely induced by the bulk criticality. Therefore, the surface critical behavior is in the ``ordinary" class. 
However, with surface couplings enhanced, the surface may order by a surface transition while the bulk is disordered.
At the bulk transition point, the ordered surface exhibits extra singularities; such a transition is in the ``extraordinary" class.
At a fine-tuned surface coupling strength, a multi-critical point occurs between the two SCBs, known as the ``special" class.

The subject of SCB has attracted numerous investigations in history\cite{Binder1983,Diehl} due to its rich and novel properties
and obtained renewed attention recently when quantum phase transitions are involved.
Zhang and Wang \cite{Zhang2017} studied the spin-1/2 Heisenberg model on a decorated square (DS) lattice and 
found a ``nonordinary" SCB realized without fine-tuning the surface coupling.  
According to the mapping between a $d$-dimensional quantum system and a $(d+1)$-dimensional classical system,  
the surface of the two-dimensional (2D) SU(2) quantum model maps onto the 2D surface of the corresponding three-dimensional (3D) O(3) 
classical system, which can not host a long-range order according to the Mermin-Wagner theorem \cite{Mermin1966}. As a result, there should be 
no SCB other than the ordinary one. Therefore, the nonordinary SCB must have a purely quantum origin. \cite{Zhang2017}
The authors attributed it to the property of the symmetry-protected topological (SPT) phase \cite{XiaoGang2009,Frank2010,XiaoGang2012};
the gapless edge state of the SPT phase complemented with the critical mode of the bulk leads to a multicritical behavior.

Later researches on simple 2D dimerized spin-1/2 Heisenberg models found similar nonordinary SCBs
on surfaces formed by dangling spins weakly coupled to the bulk, with surface critical exponents close to those of the nonordinary SCB found in 
the spin-1/2 DS Heisenberg model.\cite{Ding2018, Weber2018,Ding2021}
It was argued that the surface formed by dangling spins can be viewed as a spin-1/2 Heisenberg chain, which is
gapless due to the topological $\theta$-term of the spin-1/2 chain even if the bulk is in a trivial product state.\cite{Ding2018}
Such a gapless edge mode, together with the gapless bulk critical mode, leads to the nonordinary SCBs. 
However, this scenario was challenged by the finding that the dangling surface of the $S=1$ dimerized Heisenberg model shows similar 
nonordinary exponents. \cite{Weber2019}
Naively, the surface is a spin-1 Haldane chain formed by dangling $S=1$ spins in this case, which is gapped. 

Besides, the surface showing nonordinary SCBs of the spin-1/2 DS Heisenberg model is shown to be a dangling surface in 
a product state instead of an SPT state.\cite{Weber2018}
To check if the gapless edge state of an SPT phase can lead to similar nonordinary SCB, Zhu {\it et al.} \cite{Zhu2021}studied a 2D model of 
coupled spin-1 Haldane chains (CHC).
In the path integral representation of the spin chain, the action has a topological term that is ineffective for
integer spin-S \cite{Haldane1985}. The chain is then described by the O(3) nonlinear sigma model without the topological term.
However, this description only applies to the periodic boundary conditions.
For an integer spin-S chain with free boundaries, the action has a topological term of two spin-1/2 located at 
the ends of the chain.\cite{Abanov2017} For odd integer S, the state is an SPT protected by certain 
symmetries.\cite{Frank2010}
When integer spin-S chains are coupled to form a 2D SPT phase\cite{Wierschem2014}, it is, therefore, natural to assume 
that the spin-1/2 excitations at the ends of 
the chains form a gapless edge state, according to the Lieb-Schultz-Mattis theorem \cite{LIEB1961}.
This scenario was verified and used to explain the nonordinary SCB on the surface formed by the chain ends.\cite{Zhu2021}
We emphasize that the surface is not formed by dangling spins. Instead, the gapless edge mode is due to a genuine 
SPT phase of the model. 
On the other hand, the SCB on the surface formed by a spin-1 chain is in the ordinary class.

      \begin{figure}[htb]
	\centering
	\includegraphics[width=0.5\textwidth]{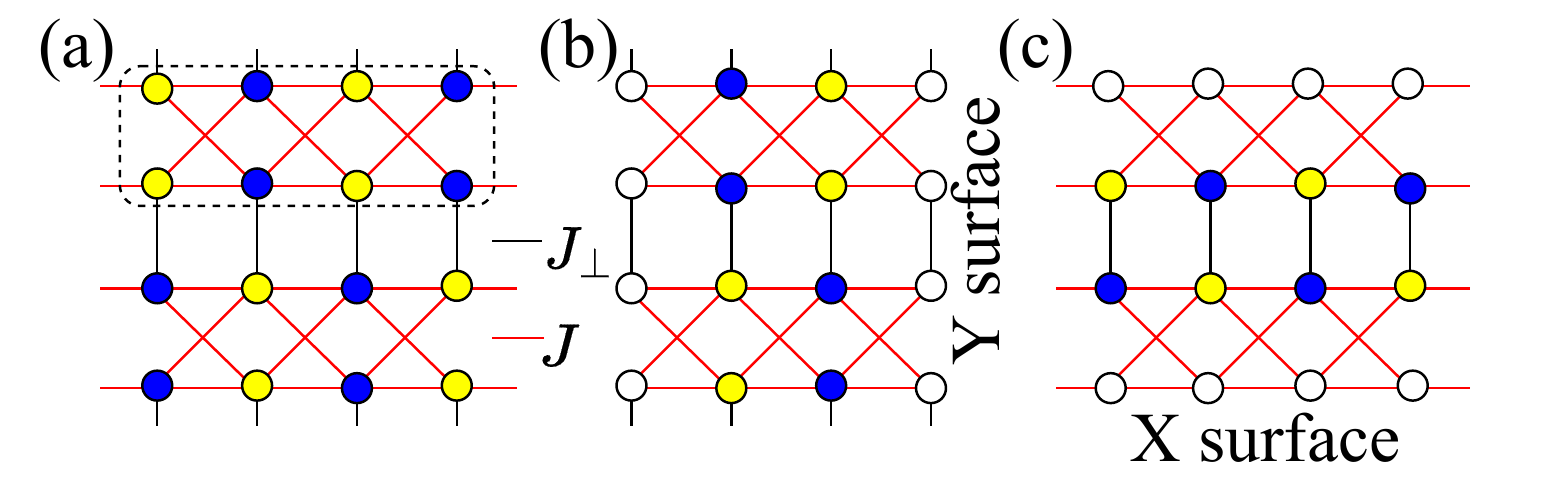}
	\caption{ The two-dimensional coupled diagonal ladders.  A diagonal ladder is shown inside the dashed rectangular box.
    The lattice is bipartite with sublattices A (yellow circles) and B (blue circles). (a) Periodic
    boundary conditions are applied in $x$ and $y$ directions. 
    (b) Periodic boundary conditions are applied in $y$ direction, while open boundaries are applied in $x$ direction to expose Y surfaces. (c) Periodic boundary conditions are used in $x$ direction, while 
    open boundary conditions are applied in $y$ direction to expose X surfaces. Open circles denote spins on the surfaces.
    The intraladder couplings $J>0$ are indicated by red lines, and interladder couplings $J_{\perp}>0$ by black lines. }
	\label{Fig:model}
    \end{figure}

     \begin{figure}[htb]
	\centering
	\includegraphics[width=0.5\textwidth]{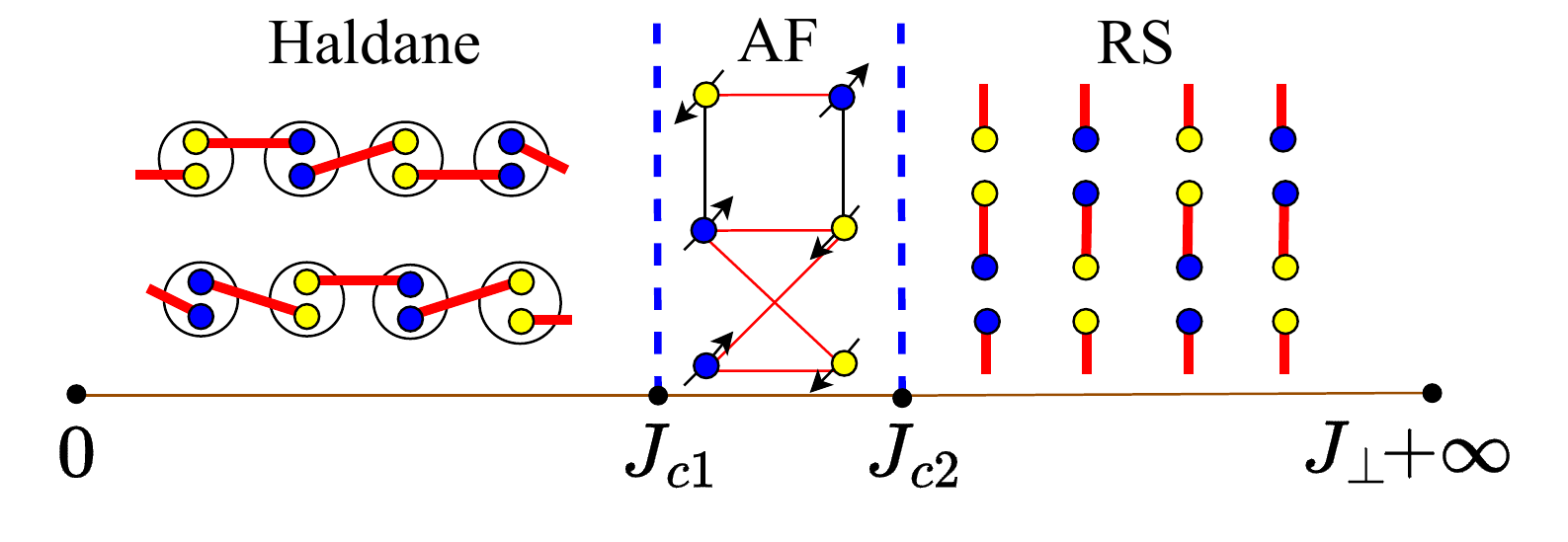}
	\caption{Phase diagram with three phases: the antiferromagnetic phase (AF), the Haldane phase, and the rung single 
	phase (RS), separated by two quantum critical points $J_{c1}$ and $J_{c2}$.
	%are realized by tuning $J_{\perp}$ and set $J$ to be unity. 
	A cartoon of a representative ground state is graphed in each phase. 
	The arrows represent the orientation of spins. Thick red lines denote spin singlets. 
	The circles in the Haldane phase indicate that two spin-1/2 form a spin-1.}
\label{Fig:diagram}
\end{figure}

Other mechanisms of such nonordinary SCB were also proposed.\cite{Metlitski2020, Jiancm2021, ParisenToldin2021,ToldinPRL2022,Meineri_scipost2022} 
In particular, an extraordinary-log SCB \cite{Metlitski2020} is proposed and proved numerically \cite{ParisenToldin2021} 
for a critical classical model in the 3D O(3) universality class. A special transition was found between 
the ordinary and the extraordinary-log transitions, with exponents close to the nonordinary exponents of the 
quantum models. It is then conjectured that those quantum models are sitting close to the special transition point by chance and showing such special
SCB at finite sizes.
\cite{ParisenToldin2021}    

In consideration of the current confusing research situation, to further investigate the origin of the nonordinary SCB in 2D quantum models,  
it would be beneficial to find/construct and study more 2D quantum SU(2) models with nonordinary SCBs existing at its 3D O(3) critical point. 
In this work, 
we construct a model that hosts an SPT phase with nontrivial gapless surface states and a simple product phase with 
gapless mode on the surfaces formed by dangling spins. Notably, the two surfaces are different. 
Two quantum critical points in the 3D O(3) universality class separates the SPT phase
and the product phase from the N\'eel phase in the middle, respectively.
We show that the two gapless edge modes induce nonordinary SCBs {\it without fine-tuning surface couplings} on different surfaces with exponents agreed 
well at different bulk critical points.

The model is constructed by coupling the spin-1/2 diagonal ladders to form a 2D lattice, as illustrated in Fig. \ref{Fig:model}.
A spin-1/2 diagonal ladder is shown in Fig. \ref{Fig:model}(a), which is the composite spin representation of a spin-1 chain,
in the sense that the low energy spectra of the two systems are identical.\cite{diagonal2000}
The ground state of the diagonal ladder is gapped and unique if periodic boundary conditions 
are applied; however, it is fourfold degenerate for open boundary conditions due to spin-1/2 degrees of freedom living on the ends of the ladder. 
In the weak coupling region $J_{\perp} \ll J$, the model should stay in the Haldane phase, which is an SPT phase with gapless edge modes at 
the surface formed (Y surfaces shown in Fig. \ref{Fig:model}(b)) by the ends of the ladders.
When the coupling $J_{\perp}$ is strong enough, the model goes to the trivial product state: rung singlet phase (RS),
which has a gapless mode on the surface formed by dangling spins (X surfaces, see Fig. \ref{Fig:model}(c)). 
Since the lattice is bipartite, when the couplings between the ladders are competitive with the couplings inside the ladders, the model should 
stay in the N\'eel phase.  
The three phases are separated by two quantum critical points(QCPs). The phase diagram is sketched in Fig. \ref{Fig:diagram}.

Using unbiased quantum Monte Carlo (QMC) simulations \cite{Sandviksusc1991,Sandvik1999}, we determine the two quantum critical points and show that
they belong to the 3D O(3) universality class regardless the nonmagnetic phase is an SPT phase or a trivial product phase.
We demonstrate that nonordinary SCBs are realized at the two bulk critical points but on different surfaces, which are 
exposed by simply cutting bonds without fine-tuning surface couplings required to reach a multicritical point in the classical models.
We study the properties of the two gapped phases, including the finite-size behavior of the string orders in the Haldane phase.
We show that the surface formed by the ladder ends is gapless in the Haldane phase, on which the nonordinary SCB is observed;
the surface along the chain direction is gapped; therefore, the SCB on it is ordinary.
However, in the RS phase, the latter surface becomes a chain formed by dangling spins and gapless. So we found nonordinary SCB on it, instead.

    The paper is organized as follows. In Sec. \ref{Sec:mm}, we describe our model and the methods used in our study.  
    Sec. \ref{sec:bulk}, we study the phase diagram and bulk quantum phase transitions of the model. The topological properties 
    of the Haldane phase and surface properties of the surfaces in two magnetic disordered phases are also studied.
    In Sec. \ref{sec:scb}, we study the surface critical behaviors of our model.  
 Finally, we conclude and discuss the mechanisms of the origins of nonordinary surface critical behavior in Sec. \ref{sec:conclsn}.

\section{Models and methods}
\label{Sec:mm}

We study the spin-1/2 Heisenberg model on a designed two dimensional bipartite lattice 
constructed by coupling diagonal ladders \cite{diagonal2000}, see Fig. \ref{Fig:model}(a). 
We will refer to the lattice as coupled diagonal ladders (CDL).
The Hamiltonian is given by
 \begin{equation}
	\begin{split}
	H &= \sum_{j=0}H_{j}+ J_{\perp}\sum_{i,j=0}\mathbf{S}_{i,2j+1}\cdot \mathbf{S}_{i,2(j+1)},
		\end{split}
	\end{equation}
	where the first sum is over the diagonal ladders with $H_{j}$ describing the $j$-th ladder written as follows 
%The spin-1/2 Heisenberg model on a diagonal ladder is described by
	\begin{equation}
	\begin{split}
	H_j &= J \sum_{l=0,1} \sum_i \mathbf{S}_{i,2j+l}\cdot \mathbf{S}_{i+1,2j+l}\\
	&+ J\sum_{i}[\mathbf{S}_{i,2j}\cdot \mathbf{S}_{i+1,2j+1}+\mathbf{S}_{i,2j+1}\cdot \mathbf{S}_{i+1,2j}],
		\end{split}
		\label{H_ladder}
	\end{equation}
where $l=0,1$ denote two legs of the $j$-th diagonal ladder, $J>0$ is  intraladder Heisenberg exchange interactions.
The second sum describes the coupling of the neighboring ladders with the interladder couplings $J_{\perp}>0$. 

We set $J$ to be unity. When $J_{\perp}$ is comparable to $J$, the model is expected in the N\'eel phase. For the limit $J_{\perp} \gg 1$, the model
is tuned into a disordered rung singlet phase (RS), a product of singlets. On the other limit, $J_{\perp} \to 0$, the model
is tuned into a gapped Haldane phase, which can be described by the AKLT state.  
These three phases are separated by two quantum critical points, as sketched in Fig. \ref{Fig:diagram}.

The lattice is bipartite; therefore, the model is free of magnetic frustration and can be studied using quantum Monte Carlo (QMC)
simulations. In this work, we use the stochastic series expansion (SSE) quantum Monte Carlo simulations with the loop 
algorithm \cite{Sandviksusc1991,Sandvik1999} to study the bulk and surface properties of the gapped phases, as well as the bulk and surface 
critical behaviors.
Periodic boundary conditions are applied along both x and y directions when the bulk phase transitions are studied.
When the surface states and surface critical behaviors are studied, periodic boundary conditions are applied along one lattice 
direction and open boundary conditions are used along the other direction to expose the surfaces, as shown in 
Fig. \ref{Fig:model}(b) and (c). The Y surfaces are obtained by cutting the lattice along the y direction. Similarly, 
we can expose the X surfaces by cutting the lattice along the x direction. Note that, for the X surfaces, we only consider the case of cutting 
$J_\perp$ bonds. 

In our simulations, we have reached linear size up to $L=128$. The inverse temperature scales as $\beta=2L$, considering the 
dynamic critical exponent $z=1$ for the two critical points. Typically $10^{8}$ Monte Carlo samples are taken for 
each coupling strength. 
 
\section{ Bulk Results}
\label{sec:bulk}
\subsection{Bulk phases and properties of associated bulk critical points}

\begin{figure}[htb]
	\centering
	\includegraphics[width=0.46\textwidth]{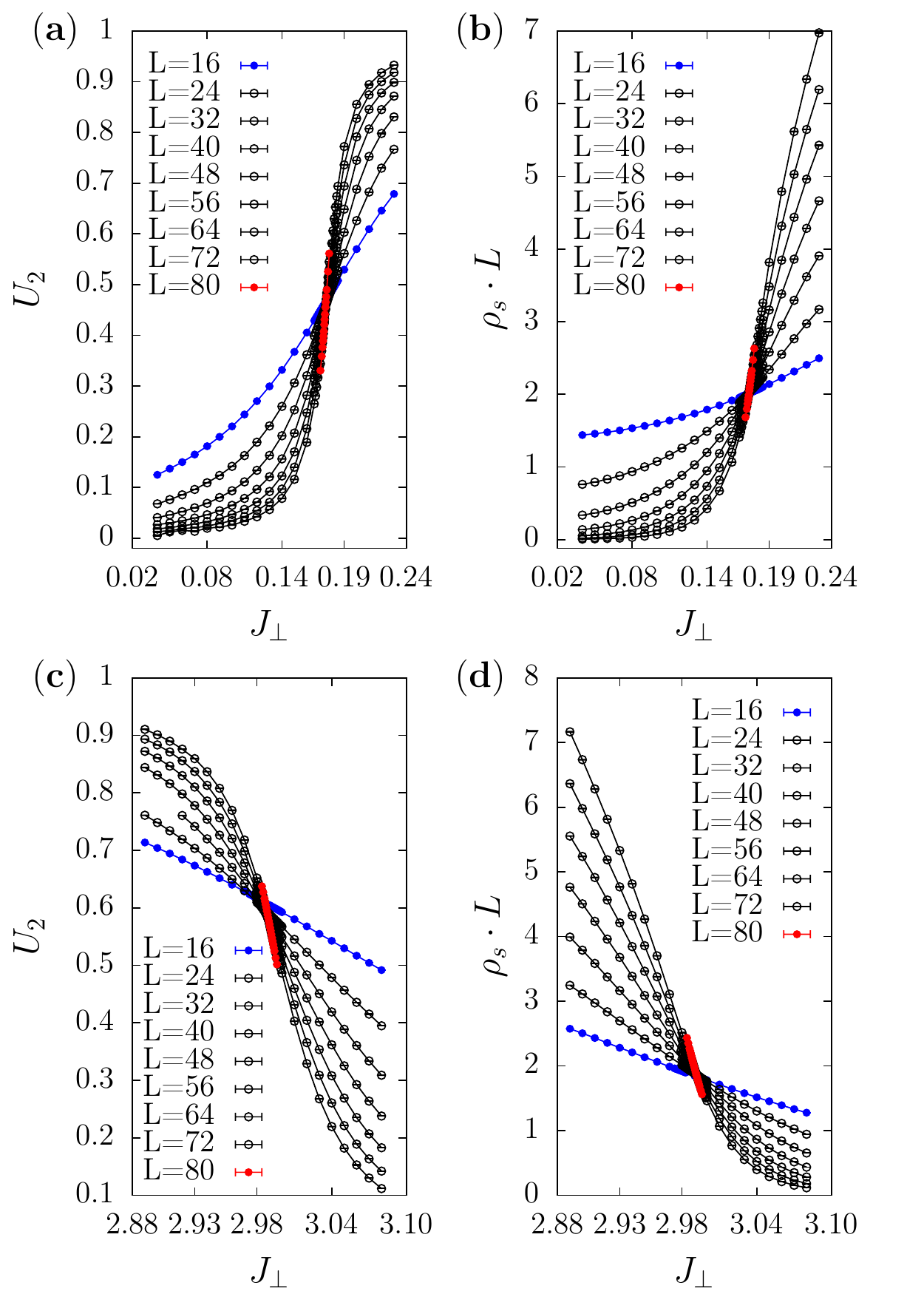}
	\caption{Binder cumulant $U_{2}$ and spin stiffness multiplied by the system size $\rho_{s}L$ versus $J_{\perp}$ 
	for different system sizes. Error bars are much smaller than the symbols. (a) and (b) shows results  near the critical point $J_{c1}$. (c) and (d) shows results near critical point $J_{c2}$ }
	\label{fig:binder}
\end{figure}

We study several physical quantities to investigate the bulk phases and related phase transitions.
The two transitions are associated with the spontaneously breaking of the spin rotational symmetry.  
The staggered magnetization is used to describe the N\'eel order,
\be
 m_s^z=\frac{1}{L^2}\sum_i \phi_{i} S_i^z,
 \ee
where the staggered phase factor $\phi_{i}=\pm 1$ according to the sublattice to which site $i$ belongs. 
The Binder cumulant $U_{2}$ \cite{Binder1981,Binder1984} is defined using $m_s^z$ 
\be
 U_{2}= \frac{5}{6}\left(3-\frac{\langle (m_{s}^{z})^{4}\rangle}{\langle (m_{s}^{z})^{2}\rangle^{2}}\right),
 \ee
which is dimensionless at the critical point.
$U_{2}$ converging to 1 with increasing system size indicates the existence of magnetic order, while tending to 
zero with increasing system size implies that the system is in the magnetic disordered phase. 

The mean spin stiffness $\rho_s$ over the $x$ and $y$ directions is also calculated. It is related to the fluctuations 
of the winding number\cite{Pollock1987,SandvikAIP}
\be
 \rho_{s}=\frac{3}{4}\langle W^2_{x}+ W^2_{y}\rangle/\beta,
 \ee
 where
  \be
  W_{a}=\frac{1}{L_a} (N^+_a-N^-_a)=0,\pm 1, \pm2, ...
 \ee
is the winding number along the $a=x, y$ direction. Here, $N^+_a$ and $N^-_a$ denotes the total number of operators 
transporting spin in the positive and negative $a$ direction, respectively.

$\rho_{s}$ is none zero if the state is magnetically ordered and goes to zero when the system is in the 
magnetically disordered phase. 
The size dependence of the spin stiffness exactly at a QCP is expected as follows \cite{Fisher1989}
\be
 \rho_{s}\sim L^{2-(d+z)},
 \ee
with $z$ the dynamic exponent and $d=2$ the dimensions of the model.  
In the case that %has an asymptotic Lorentz invariance with the dynamical critical exponent 
$z = 1$, the $\rho_{s}L$ is expected to be dimensionless at critical point. 

Figure \ref{fig:binder} plots $\rho_{s}L$ and $U_2$ as functions of $J_{\perp}$ for different system sizes.
Clearly, the model is in the antiferromagnetic ordered state when $J_{\perp}$ is between 0.17 and 3.
Since $U_2$ and $\rho_s L$ are dimensionless at a critical point, the crossings of curves for different sizes roughly indicate 
two transition points.

We adopt the standard $(L,2L)$ crossing analysis for $U_{2}$ and $\rho_{s}L$ to estimate the critical point and 
critical properties.\cite{shao_science}
For $Q=U_2$ or $\rho_s L$, 
we define the finite-size estimator of the critical points $J_{c}^{(Q)}(L)$ as the crossing point of $Q(J_{\perp})$ curves for $L$ 
and $2L$, which drifts toward the critical point $J_{c}$ in the thermodynamic limit in the following way 
\be
J_c^{(Q)}(L)= J_c +a L^{-1/\nu-\omega},
\ee
where $\nu$ is the correlation length exponent, $\omega>0$ is the effective irrelevant exponent, and $a$ is an unknown 
constant. 
At the crossing point $J_c^{(Q)}(L)$, we define the finite-size estimator of exponent $\nu$ as 
\be
\frac{1}{\nu^{(Q)}(L)}=\frac{1}{\ln 2} \ln \left(\frac{s^{(Q)}(2L)}{s^{(Q)}(L)}\right)
\ee
where $s^{(Q)}(L)$ is the slope of the curve $Q(J_\perp)$ for size $L$ at $J_c^{(Q)}(L)$.
$\nu^{(Q)}(L)$ converges to the exponent $\nu$ in the following way
\be
\nu^{(Q)}(L)=\nu + b L^{-\omega},
\ee
with $b$ an unknown constant.

For $U_2$ and $\rho_s L$, the analyses yield consistent estimates of $J_c$ and $\nu$
within error bars. The results with higher accuracy are selected as the final results. 
All the results are listed in Table \ref{ext1}. In particular, the final estimates of the 
two critical points are $J_{c1}=0.17425(3)$ and $J_{c2}=2.99046(5)$. 

\begin{ruledtabular}
\begin{table*}[!t]
	\caption{ Bulk critical properties. 	% Reduced $\chi^2$ (R-$\chi^2$) and p-value of $\chi^2$ (P-$\chi^2$) are listed below the corresponding exponents. 
	The exponents  obtained by field theory (FT) and by Monte Carlo simulations (MC) are listed for comparison. 
	}
\begin{tabular}{l c c c c  }
                                  &$J_{c}$         &  $\nu$            & $\eta$             &$y_{h}$ \\
\hline
	$J_{c1}$		            	& 0.17425(3)  &      0.707(48)     & 0.033(6)   		& 2.484(1)	 \\
\hline
	$J_{c2}$			        	&2.99046(7)   & 0.705(6)           &0.0324(34)           &2.483(1)   \\
\hline
	FT\cite{Guida1998}		    	&    		  & 0.7073(35)   		 	&0.0355(25)    &   \\
	MC\cite{Hasenbusch2011}			&    			& 0.7117(5)   		 	&0.0378(3)    &   \\
\end{tabular}
\label{ext1}
\end{table*}
\end{ruledtabular}

To further determine the universal properties of the two critical points, we calculate the static spin structure factor and the spin 
correlation at the longest distance in a finite system at the two estimated critical points $J_{c1}$ and $J_{c2}$. 
The two quantities are defined based on the spin correlation function
\begin{equation}
 C(\mathbf{r}_{ij})=\langle S^z_i S^z_j \rangle,
\end{equation}
%at half of the lattice size $\mathbf{r}$=(L/2,L/2) $C(L/2,L/2)$ and
where $\mathbf{r}_{ij}$ is the vector from site $i$ to $j$. 
The static spin structure factor at wave vector $(\pi,\pi)$ is defined as follows
 \begin{equation}
 S(\pi,\pi)=\sum_{\mathbf{r}} \epsilon_{ij} C(\mathbf{r}_{ij}),
\label{spipi}
\end{equation}
 where $\epsilon_{ij}=\pm 1$, depending on whether $i$ and $j$ belong to the same sublattice.
The spin correlation function of half lattice size $C(L/2,L/2)$ averages $C(\mathbf{r}_{ij})$ between two spins $i$ and $j$ 
at the longest distance $\mathbf{r}_{ij}=(L/2,L/2)$.

$S(\pi,\pi)$  and $C(L/2,L/2)$ are used
to extract the scaling  dimension $y_{h}$ of the staggered magnetic field $h$ and the anomalous dimension  $\eta$. 
At QCP, $S(\pi,\pi)$  and $C(L/2,L/2)$
satisfy the following finite size scaling forms
\begin{equation}
 S(\pi,\pi)/L^2\sim L^{-2(d+z-y_{h})}(1+bL^{-\omega}),
\label{spi}
\end{equation}
and
\begin{equation}
 C(L/2,L/2)\sim L^{-(d+z-2+\eta)}(1+bL^{-\omega}),
\label{cb}
\end{equation}
respectively, in which $d=2$ is the spatial dimension, $z=1$ is the dynamical critical exponent, and $\omega$ the effective correction to scaling exponent.
%and \note{$\omega=1$  yields good fitting to all critical exponents}. 
The two exponents $y_h$ and $\eta$ are not independent and are expected to obey the following scaling relation
\begin{equation}
\eta=d+z+2-2 y_{h}.
\label{scalingb}
\end{equation}

\begin{figure}[htb]
	\centering
	\includegraphics[width=0.46\textwidth]{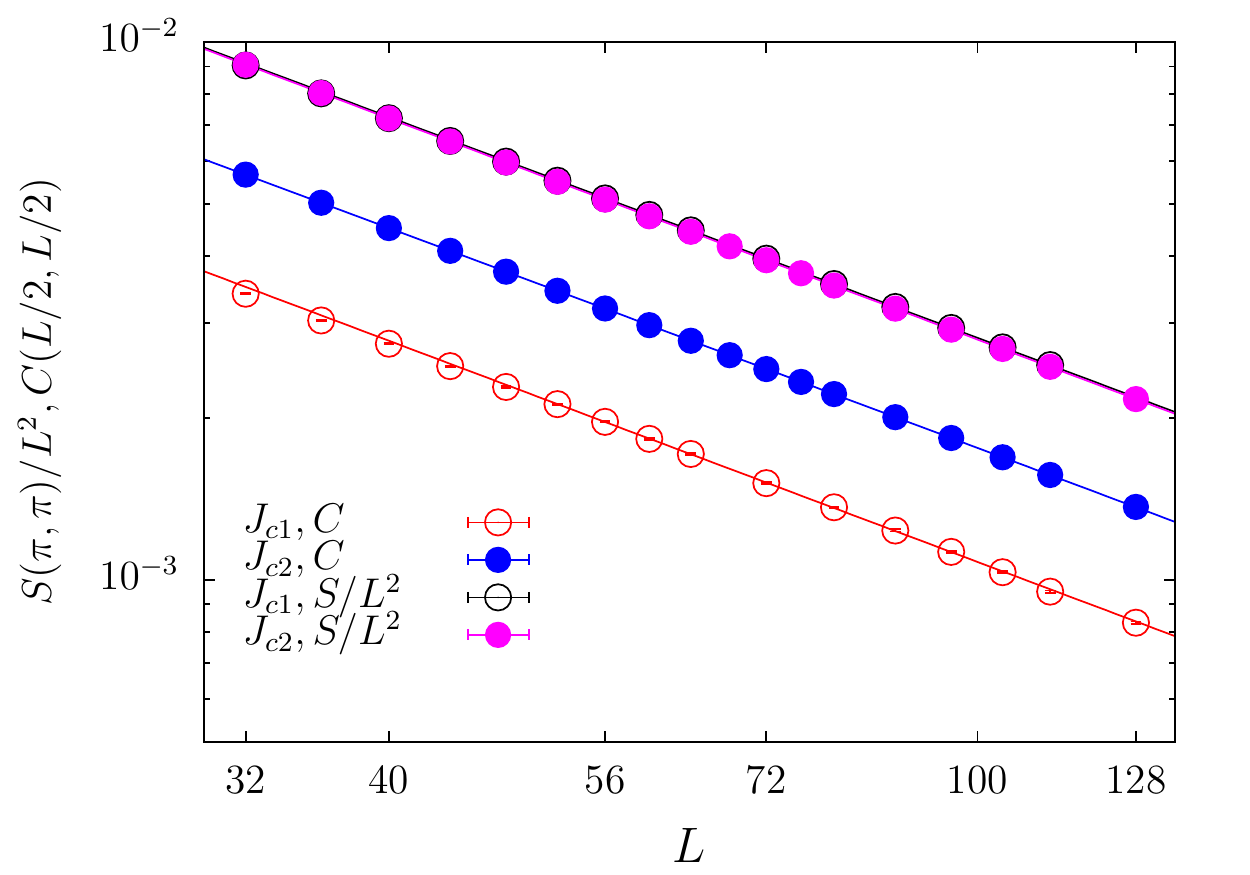}
	\caption{
	$C(L/2,L/2)$ and $S(\pi,\pi)/L^2$ versus system size $L$ at quantum critical points $J_{c1}$ and $J_{c2}$ on a 
	log-log scale. The symbols of $S/L^2$ at $J_{c1}$ are covered by those at $J_{c2}$. Error bars are much smaller than the symbols.}
	\label{fig:bulkcandm}
\end{figure}

The numerical results of $ S(\pi,\pi)/L^2$ and $C(L/2,L/2)$ as functions of system size $L$ at two critical points 
are shown in Fig.~\ref{fig:bulkcandm}.
We fit the data of $S(\pi,\pi)/L^2$ and $C(L/2,L/2)$ according to Eqs. (\ref{spi}) and (\ref{cb}), respectively, and find the critical
exponents $y_{h}$ and $\eta$, as listed in Table \ref{ext1}. 
The obtained $y_{h}$ and $\eta$ satisfy the scaling relations Eq. (\ref{scalingb}).

Comparing with the best known exponents of the 3D $O(3)$ universality class\cite{Guida1998,Hasenbusch2011}, we conclude that both 
critical points belong to the 3D $O(3)$ universality class.  
This also shows that the topological order does not change the universality class of the bulk phase transition.

\subsection{Properties of the Haldane phase and its surface states}
\label{subsec:Haldane}

\begin{figure}[htb]
	\centering
	\includegraphics[width=0.46\textwidth]{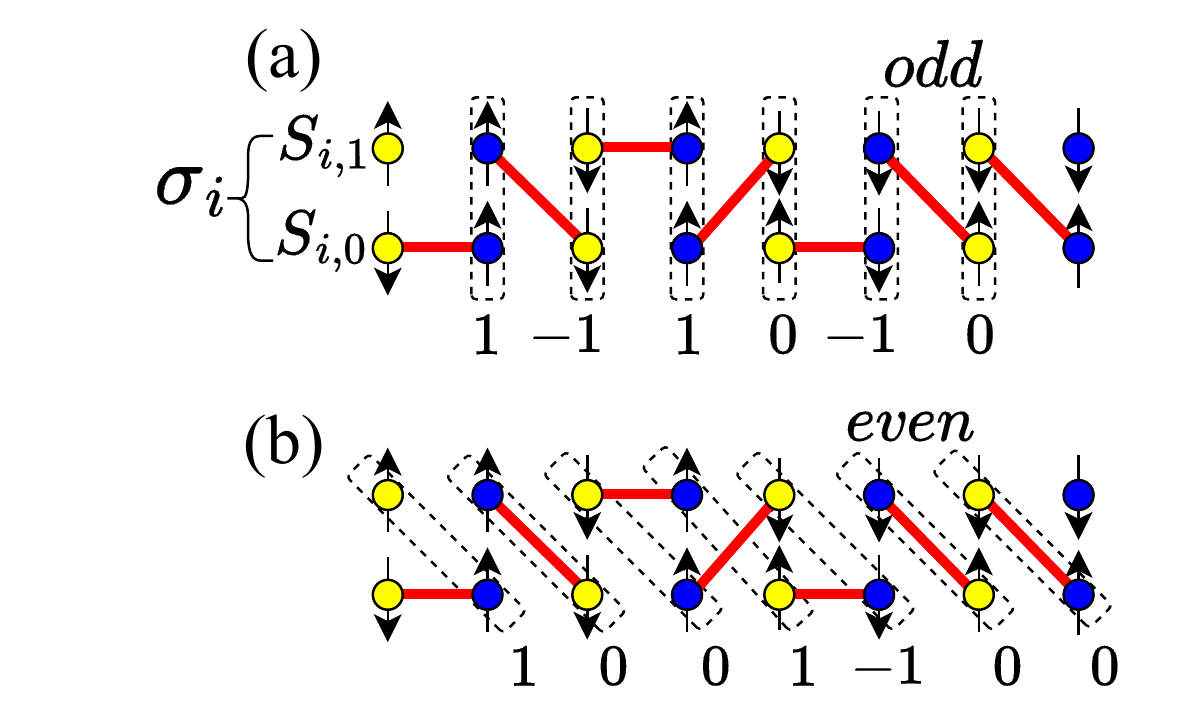}
	\caption{
	A representation of VB ground state with two spin-1/2 localized at the ends of the ladder and
	a spin configuration matches the VB state. %is also graphed. 
	(a) Dashed boxes encircle rungs: $S_i^z=S^z_{i,0}+S^z_{i,1}$, used to calculate the string order parameter $\mathcal{S}_{\rm odd}$. After all sites with
	$S_i^z=0$ removed, the remaining sites show N\'eel order.
	(b) Dashed boxes encircle diagonals: $S_i^z=S^z_{i+1,0}+S^z_{i,1}$, used to  
	calculate the string order parameter $\mathcal{S}_{\rm even}$.
	After all sites with $S_i^z=0$ removed, the remaining sites do not show N\'eel order.
	}
	\label{Fig:composite}
\end{figure}

 The spin-1/2 Heisenberg diagonal ladder is shown as the composite spin representation of a spin-1 chain\cite{diagonal2000} 
 by representing the spin-1 operator $\boldsymbol{\sigma}_{i}$ of the chain as the sum of two spin-1/2 operators $\boldsymbol{\sigma}_{i}=\mathbf{S}_{i,0}+\mathbf{S}_{i,1}$ on two legs, as illustrated in Fig. \ref{Fig:composite}.
The low-energy spectrum of the ladder is identical to that of the spin-1 chain. 
When periodic boundary conditions are applied, all spins are bound to form valence bonds and the ground state is unique.
The ground state of this ladder can be well described by the AKLT state\cite{AKLT}, which is a short-ranged 
valence-bond (VB) state.
A typical configuration  is shown in Fig. \ref{Fig:diagram}. The Haldane gap is related to the energy needed to 
break a valence bond. More importantly, with open boundaries, the ground states have two spin-1/2 spins localized at the ends of the ladder.
This is evident in the diagonal ladders as shown in Fig. \ref{Fig:composite}. 
The ladder is in a Haldane phase with symmetry protected topological order.

The Haldane phase with symmetry protected topological order is characterized by a nonlocal string order, which
is evident when the $z$ component of the spins on the same rung are summed. 
The total $S_i^z=S^z_{i,0}+S^z_{i,1}$ can take the values of 1,0, -1. When the sites with $S_i^z=0$ are removed, 
the remaining sites have a N\'eel order, which means a string order, as shown in Fig. \ref{Fig:composite}(a). This order 
can be described by the following string order parameter: \cite{Nijs1989, diagonal2000}
	\begin{equation}
	\begin{split}
	\mathcal{S}_{\rm odd}(i, j) &= \langle (S_{i,0}^z+S_{i,1}^z)\\
	& \exp( i \pi \sum_{k=i+1}^{j-1}(S_{k,0}^z+S_{k,1}^z))(S_{j,0}^z+S_{j,1}^z)\rangle.
	\label{s}
		\end{split}
	\end{equation}
The name $\mathcal{S}_{\rm odd}$ comes from the topology of the VB's in the diagonal ladder \cite{diagonal2000}, which 
is determined by the parity of the number of VB's crossing an arbitrary vertical line. Note that the VB state shown
in Fig. \ref{Fig:composite} is odd.

	Other ladders in the Haldane phase may show a topologically distinct string order, which can be defined as
\begin{equation}
	\begin{split}
	\mathcal{S}_{\rm even}(i, j) &= \langle (S_{i+1,0}^z+S_{i,1}^z) \\
\label{s1}
	& \exp( i \pi \sum_{k=i+1}^{j-1}(S_{k+1,0}^z+S_{k,1}^z))(S_{j+1,0}^z+S_{j,1}^z)\rangle,
		\end{split}
	\end{equation}
which is nonzero when the VB configuration is even,i.e., an even number of VB's crossing an arbitrary vertical
line\cite{diagonal2000}. 
In the case of diagonal ladder, as shown in Fig. \ref{Fig:composite}(b), when the $z$ component of the spins along the plaquette diagonals are 
summed, there is no such a string order. 
Apparently, the parity of the VB ground state is intimately related to the type of string order.

\begin{figure}[htb]
	\centering
	\includegraphics[width=0.46\textwidth]{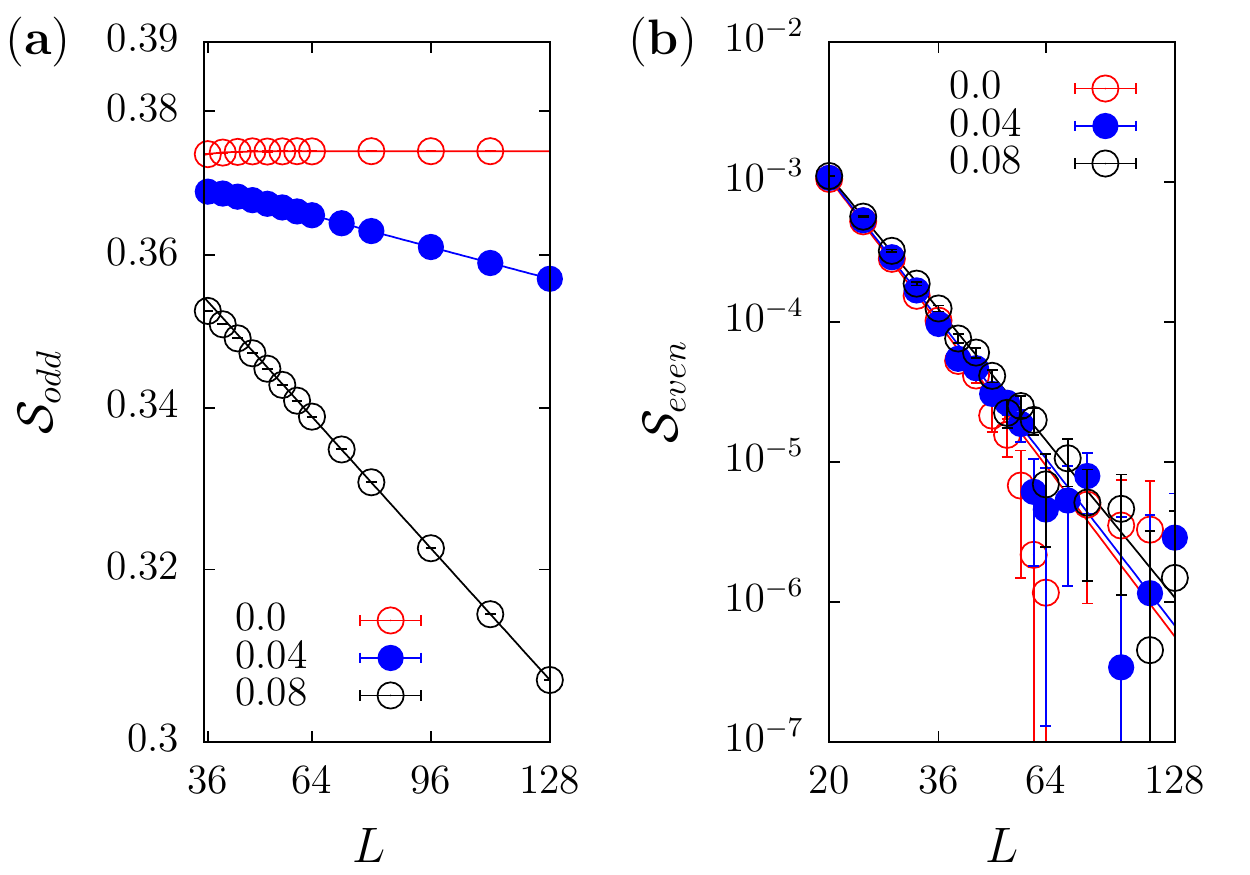}
	\caption{String order parameters $\mathcal{S}_{\rm odd}(L/2)$ (a) and $\mathcal{S}_{\rm even}(L/2)$ (b)  of the 2D 
	coupled diagonal ladders at different interladder coupling $J_{\perp}$ in the SPT Haldane phase. 
%	(c) and (d) are in the trivial RS phase. 
 (a) on a linear-log scale. (b) on a log-log scale. }
	\label{Fig:stringorder}
\end{figure}

The finite value of a string order parameter at the limit $|i-j| \to \infty$  characterizes a stable topological order 
in the thermodynamic limit. 
In the simulations of a system of size $L$ with periodic boundaries, we calculate $\mathcal{S}_{\rm odd}(L/2)$ 
($\mathcal{S}_{\rm even}(L/2)$) by averaging $\mathcal{S}_{\rm odd}(i,j)$ ($\mathcal{S}_{\rm even}(i,j)$ )
at the maximum available distance $|i-j|=L/2$ along an individual ladder.
As shown in Fig. \ref{Fig:stringorder}(a) and (b)), in the case $J_{\perp}=0$, $S_{\rm odd}(L/2)$ is finite, 
$S_{\rm even}(L/2)$ vanishes  when $L$ goes infinity,
as expected for a diagonal ladder. We find $\mathcal{S}_{\rm odd}$ converges to 0.374325(7).

Now consider that the diagonal ladders are coupled to form a 2D lattice by interladder couplings $J_{\perp}>0$.
Due to the Haldane gap, the properties of the ground state are robust against a weak higher-dimensional coupling 
between ladders.
There is no phase transition  when $J_{\perp}$ is turned on to finite values less than $J_{c1}$. This means that 
the model has a gapped Haldane phase that is adiabatically connected to the AKLT states of diagonal ladders. 
However, it was predicted theoretically that the string order of the coupled spin-1 Haldane chains 
is not stable and decays 
exponentially for arbitrarily weak interchain coupling \cite{Anfuso2007}. 
This prediction has been verified numerically recently in the 
2D CHC model\cite{Zhu2021} for sufficient large system sizes.

We obtain similar results in the SPT Haldane phase of the current model. We find that the string order parameter 
$\mathcal{S}_{\rm odd}(L/2)$ decays exponentially with $L$, but much slower than the decay of string order parameter 
in the CHC model at the same inter-ladder/inter-chain couplings. The numerical results are shown in 
Fig. \ref{Fig:stringorder}(a). Fitting according to the following formula \cite{Anfuso2007}
\begin{equation}
    \mathcal{S}_{\rm odd}(L/2)\sim \exp{(-\alpha L/2)},
\end{equation}
we find $\alpha=0.0007437(8)$ for $J_\perp=0.04$, and $\alpha=0.003089(1)$ for $J_\perp=0.08$. 

We also calculated $\mathcal{S}_{\rm even}(L/2)$ inside the Haldane phase. 
As expected, the even string order parameter values are much smaller than the odd one.
However, interestingly, we find that $\mathcal{S}_{\rm even}(L/2)$ decays algebraically with system size, as
shown in Fig. \ref{Fig:stringorder}(b). %This confuses us and we hope that some one can do it theoretically. 
We have tried to fit the data using the following scaling form
\begin{equation}
    \mathcal{S}_{\rm even}(L/2) \sim L^{-\beta},
\end{equation}
and find $\beta=4.2(1)$ for a single ladder, $\beta=3.97(5)$ at $J_\perp=0.04$, and $\beta=3.74(3)$ at $J_\perp=0.08$.

	\begin{figure}[htb]
	\centering
	\includegraphics[width=0.46\textwidth]{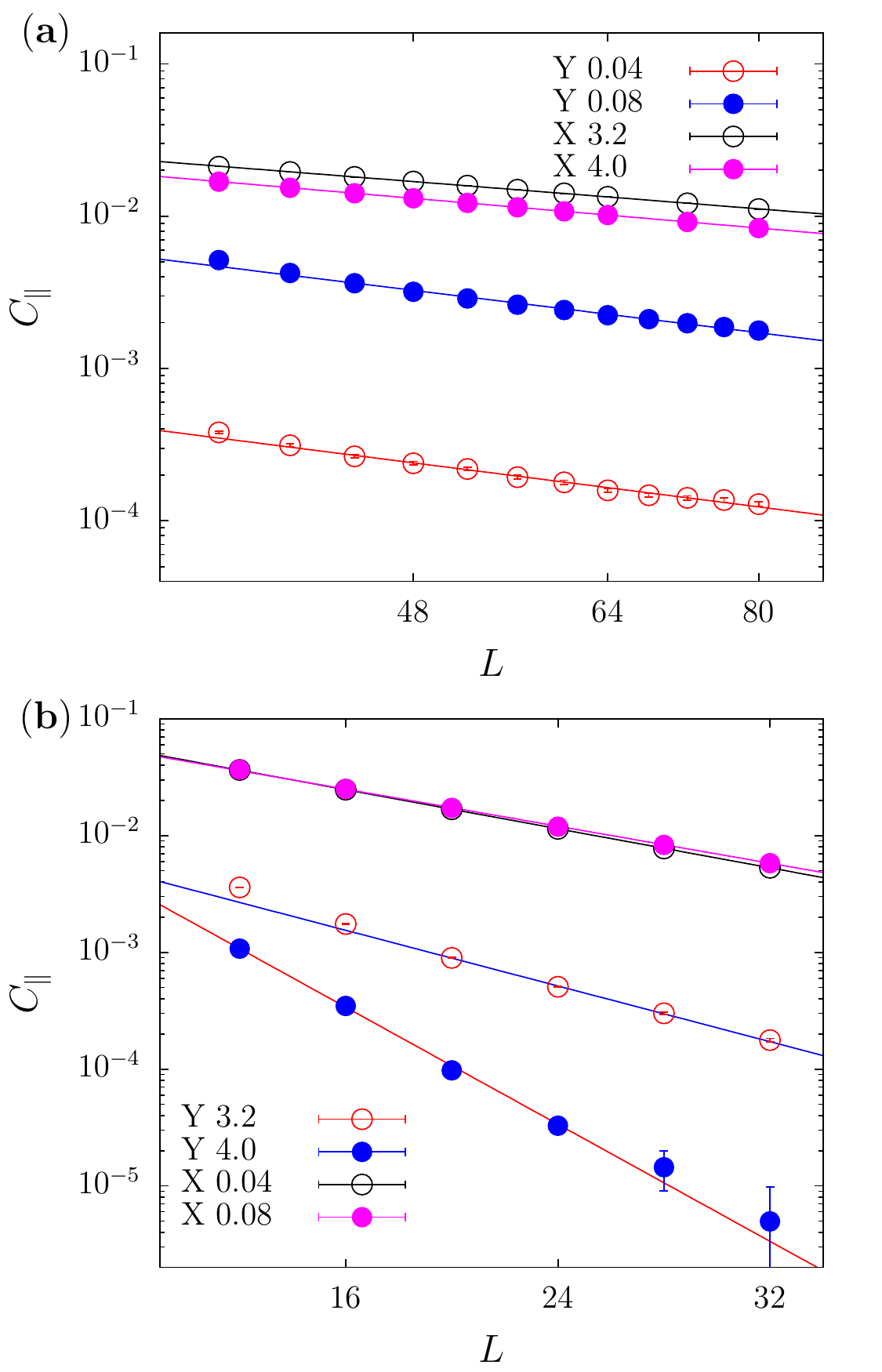}
	\caption{
	Surface correlation $C_{\parallel}(L/2)$ vs. system size $L$.
	(a) Y surface in the Haldane phase ($J_{\perp}=0.04,0.08$) and X surface in the RS phase ($J_{\perp}=3.2,4.0$). 
	The plot is on a log-log scale. Algebraically decaying with $L$ is seen, showing gapless surface states. 
		(b) Y surface in the RS phase and X surface in the Haldane phase.  The plot is set on a linear-log scale. Exponentially decaying with $L$ is observed, meaning the surface states are gapped.
		}
	\label{Fig:surfacegapbulk}
\end{figure}

%However, it is also predicted that the the edge state of the 2D SPT state is stable against dimerization. \cite{Pollmann2012}.
However, the hallmark of the SPT phase is not the string order, but the presence of nontrivial
surface states that are gapless or degenerate.
Our model is spatial anisotropy, we here consider two different surfaces, the Y surface and the X surface, 
exposed by cutting the lattice, see Fig. \ref{Fig:model}.
%One is the Y surface (Fig. \ref{Fig:model}(b)) which is exposed by cutting horrizontal $J_x$ bonds along the y direction, the 
%other one is called the X surface (Fig. \ref{Fig:model}(c), 
%which is formed by cutting $J_{\perp}$ bonds along the x direction.
To study the surface states, we calculate the surface parallel correlation $C_\parallel(L/2)$ which averages 
$C(\mathbf{r}_{ij})$ between two surface spins $i$ and $j$ at the longest distance $L/2$.

The results for the Y surface at two couplings $J_{\perp}$ in the SPT phase are plotted in 
Fig. \ref{Fig:surfacegapbulk}(a).
We see that $C_\parallel(L/2)$ decays with system size $L$ in a power law, 
\begin{equation}
    C_\parallel(L/2) \sim  L^p
    \label{gapless_surface}
\end{equation}
with $p=1.30(4)$ at $J_\perp=0.04$ and $p=1.12(2)$ at $J_\perp=0.08$,
meaning that the surface states are gapless. 
This is easy to understand from the AKLT state shown in Fig. \ref{Fig:composite}(a). With open boundaries, each ladder carries a spin-1/2 excitation 
at each end.  
An effective spin-1/2 antiferromagnetic Heisenberg (AFH) chain 
is formed at the Y surface by coupling these spins, which is gapless according to the Lieb-Schultz-Mattis theorem\cite{LIEB1961}. 

Meanwhile, the results of $C_\parallel(L/2)$ for the X surface at the same two couplings in the SPT phase have completely 
different finite-size behavior, as shown in Fig. \ref{Fig:surfacegapbulk}(b). The data can be fitted using straight lines 
on a linear-log scale, meaning the correlation decay exponentially. 
Fitting the curves with 
\begin{equation}
    C_\parallel(L/2) \sim \exp{(-L/a)},
    \label{surface_gap}
\end{equation}
we obtain $a=10.56(2)$ at $J_\perp=0.04$ and $a=11.49(3)$ at $J_\perp =0.08$.
The surface states on the X surface are gapped, because the X surface is a gapped diagonal ladder, in 
its AKLT state, weakly coupled to the bulk.

\subsection{Surface states of the trivial rung singlets product phase}
\label{subsec:rs}
Now we move to the RS phase with $J_{\perp} > J_{c2}$.
The nature of this trivial disordered phase can be understood by examining the limit $J_{\perp}\to \infty$, at which 
the lattice reduces to disjoint bonds, see Fig. \ref{Fig:diagram}. The ground state is the direct product 
of these rung singlets. 

%The 2D system can also be viewed as antiferromagnetic(AF) ladders coupled by diagonal couplings $J$.
%The ground states of these AF ladders are well described by resonating valence bond (RVB) state, however, there is no 
%free spin-1/2's at the ends of the ladder when open boundaries applied. As a result, the gapped phase of the coupled 
%AF ladders is not an SPT phase. 

We have calculated $C_\parallel(L/2)$ along the Y surface. The results are plotted in Fig. \ref{Fig:surfacegapbulk}(b). We 
see the correlations at $J_{\perp}=3.2$ and $J_{\perp}=4.0$ decaying exponentially. Fitting data according
to Eq. (\ref{surface_gap}), we find $a=7.2(2)$ and $3.46(6)$, respectively, indicating the Y surface states are gapped.
Apparently, the surface can be understood as sitting in a state adiabatically connected to a product state of dimers.

However, the X surface can be considered as a chain of dangling spins, weakly coupled to the bulk, in the
RS phase. Thus, we expect the surface states are gapless, forming a spin-1/2 Heisenberg chain. This 
is proven by our numerical results. As shown in Fig. \ref{Fig:surfacegapbulk}(a),
the correlation $C_\parallel(L/2)$ along the X surface at $J_\perp=3.2$ and $4.0$ decay in a 
power law. Fitting according to Eq. (\ref{gapless_surface}), we obtain the power $p=0.827(2)$ and $0.886(2)$, respectively.

\section{ Surface critical behaviors}
\label{sec:scb}
%With open boundaries, 1D AKLT state carries spin-1/2 topology excitation
%at the end of ladder, as shown intuitively in Fig. \ref{Fig:diagram}. 
%For 2D AKLT state, these ending spins form an efficient gapless 
%AFH chain, as studied in Sec. \ref{sec:bulk}. 

We now study the surface critical behaviors on the X and Y surfaces at the two bulk critical points, respectively.
Besides the surface correlation $C_\parallel(L/2)$, we also calculate another spin-spin correlation $C_\perp(L/2)$ and 
the surface staggered magnetic susceptibility $\chi_{s1}$ with respect to the surface field $h_1$. 

$C_\perp(L/2)$ averages $C(\mathbf{r}_{ij})$ between spin $i$ fixed on the surface and spin $j$
located at the center of the bulk, with $\mathbf{r}_{ij}$ perpendicular to the surface with $|j-i|=L/2$.

$\chi_{s1}$ can be calculated through the Kubo formula\cite{Sandviksusc1991}
 \begin{equation}
    \chi_{s1}=\frac{\partial\langle m^z_{s1}\rangle}{\partial h_{1}}= L\int_{0}^{\beta}d\tau\langle m^z_{s1}(\tau)m^z_{s1}(0)\rangle,
    \label{eq2}
\end{equation}
where $m^z_{s1}$ is the staggered surface magnetization defined as follows
\be
m_{s1}^z =\sum_{i \in {\rm surface}} \phi_i S_i^z,
\ee
where the summation is restricted on the surface, $\phi_i=\pm 1$ depending on the sublattice to which $i$ belongs.

At bulk critical points, the finite-size scaling behavior of the two correlations is characterized by two 
anomalous dimensions $\eta_\parallel$ and $\eta_\perp$, respectively;
\begin{equation}
C_{\parallel}(L/2) \sim L^{-(d+z-2+\eta_{\parallel})}(1+b_1 L^{-\omega}),
\label{cs1}
\end{equation}
and
\begin{equation}
C_{\perp}(L/2) \sim L^{-(d+z-2+\eta_{\perp})}(1+b_2 L^{-\omega}),
\label{cs2}
\end{equation}
where $\omega$  is the effective exponent of corrections to scaling, $b_1$ and $b_2$ are unknown constants.
The susceptibility $\chi_{s1}$ has the following scaling form:
\begin{equation}
\chi_{s1}\sim L^{-(d+z-1-2y_{h1})}(1+bL^{-\omega}),
\label{chis}
\end{equation}
where $y_{h1}$ is the scaling dimension of the surface field $h_1$, 
$\omega$ the effective exponent of corrections to scaling, and $b$ an unknown constant.
For our model $d=2$ and $z=1$.
$\omega=1$ yields good fitting for all critical exponents.

The three exponents $y_{h1}$, $\eta_{\parallel}$, and $\eta_{\perp}$ are related through the following 
scaling relations:\cite{Diehl}
 \begin{equation}
2\eta_\perp = \eta_\parallel +\eta  
\label{scalings1} 
\end{equation}
and
\begin{equation}
\eta_\parallel = d+z - 2y_{h1},
\label{scalings2}
\end{equation} 
with $\eta$ the anomalous magnetic scaling dimension of the bulk critical point in the $d+z$ spacetime.

In the remainder of this section, we use these physical quantities to examine SCBs. Two ordinary and two 
nonordinary SCBs on different surfaces are found. All the surface critical exponents obtained by various fits \cite{PYoung} to 
Eqs. (\ref{cs1}), (\ref{cs2}), and (\ref{chis}) 
%as well as the reduced $\chi^2$ and p-value of $\chi^2$ \cite{PYoung} in this work 
are listed in Tab. \ref{tab:scb}. 
For the reader's convenience, the surface critical exponents of other models are listed in Tab. \ref{exp3}. 
% (The sources of information are mainly referred to \cite{Zhu2021}).

\begin{ruledtabular}
\begin{table}[htb]
\caption{Surface critical exponents at different surface configurations. 
}
\begin{tabular}{l c c c }
 Configuration 	   	     & $y_{h1}$  & $\eta_{\parallel}$
& $\eta_{\perp}$	 \\
 	   	
\hline
	Y-c1				&1.756(3) &-0.511(2) & -0.237(2) \\
\hline
	Y-c2	  &0.852(46)  &1.318(31) &0.682(9) \\
\hline
	X-c1	  &0.82(1)  &1.36(6) &0.69(3) \\
\hline
	X-c2	  &1.780(2)  &-0.56(1) &-0.259(3) \\
\end{tabular}
\label{tab:scb}
\end{table}
\end{ruledtabular}

%\subsection{Ordinary transition}
\subsection{Surface critical behaviors at $J_{c2}$}
%We first consider Y surface obtained  by cutting the lattice along the y direction at QCP $J_{c2}$ (denoted by Y-c2) (see 
%Fig. \ref{Fig:model}(b)). 
%The surface states remain gapped ( see Fig. \ref{Fig:surfacegapbulk}(b) ), in the trivial bulk disordered phase adjacent to $J_{c2}$. 
%Thus the long-range correlation and the singularities in its physical quantities are purely induced by the bulk, at $J_{c2}$. 
%This is the case of an “ordinary transition”.

We first study the surface critical behaviors associated with the bulk critical point $J_{c2}$ separating the 
N\'eel ordered phase from the RS phase. 

We start with checking the SCB on the Y surface, referred to as ``Y-c2''.
The numerical result of $\chi_{s1}$ as a function of size $L$ is graphed in  Fig. \ref{fig:surface}(a), and the results 
of $C_{\parallel}(L/2)$ and $C_{\perp}(L/2)$ as functions of $L$ are plotted in Fig. \ref{fig:surface}(b).

We fit the data of  $C_{\parallel}(L/2)$ and $C_{\perp}(L/2)$ according to Eqs. (\ref{cs1}) and (\ref{cs2}) and 
find statistically sound estimates of $\eta_{\parallel}=1.318(31)$ and $\eta_{\perp}=0.682(6)$.

The finite-size scaling form Eq. (\ref{chis}) supplemented with a constant $c$ as non-singular 
contribution, i.e., 
\begin{equation}
    \chi_{s1}=c+ a L^{-(2-2y_{h1})}(1+bL^{-\omega}),
\end{equation} 
is used to fit the data of $\chi_{s1}$. The estimate of $y_{h1}$ is 0.852(34), with $\omega$ setting to 1.  

These surface exponents are listed in Tab. \ref{tab:scb}.
They obey the scaling relations in Eqs. (\ref{scalings1}) and  
(\ref{scalings2}), and agree well with the universal class of the ordinary transition associated with the 
3D O(3) universality class found in various classical and quantum phase transitions
(see Table \ref{exp3}). This behavior is expected since the surface state on Y surface in the RS phase is gapped,
as shown in Sec.\ref{subsec:rs}.

We then check the SCBs at critical point $J_{c2}$ on the X surface, referred to as ``X-c2''.
This is the case that the surface is made up of dangling spins.

The numerical result of $\chi_{s1}$ as a function of size $L$ is graphed in  Fig. \ref{fig:surface}(a), and
the results of $C_{\parallel}(L/2)$ and $C_{\perp}(L/2)$ as functions of $L$ are shown in Fig. \ref{fig:surface}(b).
Data fitting according to Eqs. (\ref{cs1}), (\ref{cs2}), and (\ref{chis}) finds statistically sound estimations 
$\eta_\parallel=-0.560(8), \eta_\perp=-0.259(4)$, and 
$y_{h1}=1.780(2)$, satisfying the scaling relations Eqs. (\ref{scalings1}) and (\ref{scalings2}).

The three exponents are also listed in Tab.\ref{tab:scb}.
They are consistent or very close to the nonordinary SCBs found in the quantum critical points of the 
3D O(3) universality class \cite{Zhang2017, Ding2018, Weber2018, Zhu2021}. 
This result supports the scenario that 
nonordinary SCB can be induced by the gapless surface mode on the dangling spin-1/2 surface, as explored
in Sec.\ref{subsec:rs}.

%\subsection{Nonordinary transition}
\subsection{Surface critical behaviors at $J_{c1}$}

We then study the SCBs associated with the bulk critical point $J_{c1}$ at which the SPT Haldane phase transfers to 
the O(3) symmetry broken N\'eel phase.

We first study the Y surface with associated SCB denoted by ``Y-c1". 
The surface does not consist of dangling spins. 
However, we have shown in Sec.\ref{subsec:Haldane} that the state of the Y surface in the gapped SPT Haldane phase 
is gapless due to the spin-1/2 excitations located at the ends of each diagonal ladder.

\begin{figure}[htb]
	\centering
	\includegraphics[width=0.46\textwidth]{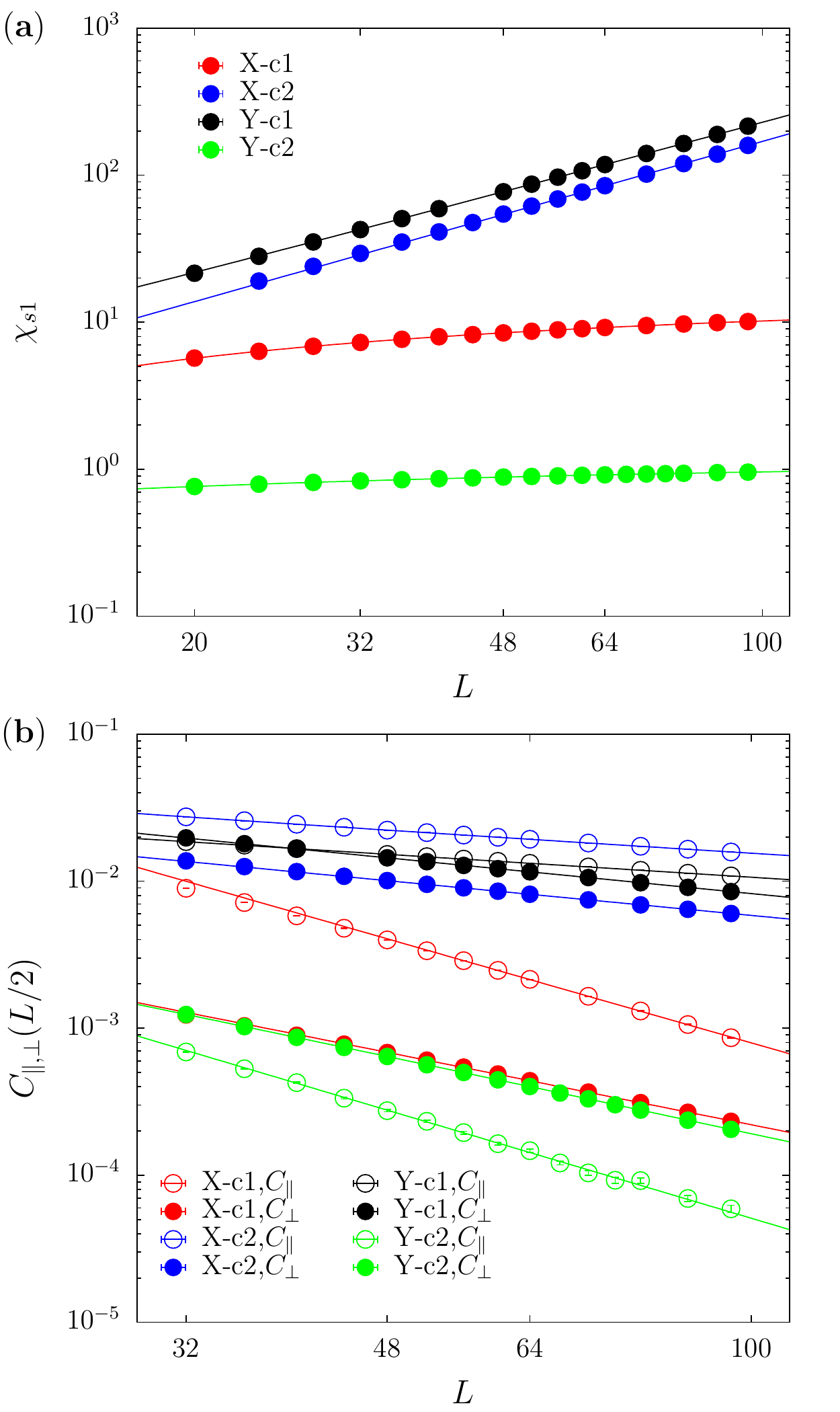}
	\caption{Surface staggered magnetic susceptibility $\chi_{s1}$ (a) 
and the correlations $C_{\parallel}(L/2)$ and $C_{\perp}(L/2)$ (b) versus system size $L$ 
for four surface configurations: X-c1, X-c2, Y-c1, and Y-c2. The plots are on log-log scale.}
	\label{fig:surface}
\end{figure}

The numerical results of $C_\parallel(L/2)$, $C_\perp(L/2)$, and $\chi_{s1}(L)$ as functions of $L$ are plotted in Fig. \ref{fig:surface}. 
Apparently, the SCBs on the Y surface (Y-c1) are similar to those of X-c2 at $J_{c2}$. 

%For the Y-c1 surface, 
We fit the data of $\chi_{s1}$, $C_{\parallel}(L/2)$, and $C_{\perp}(L/2)$ according to
Eqs. (\ref{chis}), (\ref{cs1}), and (\ref{cs2}), respectively, and obtain statistically sound results 
$y_{h1}=1.756(3)$, $\eta_{\parallel}=-0.511(2)$ and $\eta_{\perp}=-0.237(2)$, as listed in Tab. \ref{tab:scb}.
These exponents satisfy the scaling relations Eqs. (\ref{scalings1}) and (\ref{scalings2}), and 
are consistent with or very close to the nonordinary SCBs found on the X surface at $J_{c2}$, as well as 
those nonordinary SCBs found at other quantum critical points of the 3D O(3) universality
class \cite{Zhang2017, Ding2018, Weber2018, Zhu2021}.

Finally, we check the SCBs on the X surface, referred to as ``X-c1''. The surface is gapped in the Haldane phase, as
shown in Sec.\ref{subsec:Haldane}, as a result, the surface transition must belong to the ordinary class.
Our numerical results verified that the SCBs are of the ordinary type. 

The numerical results of $C_\parallel(L/2)$, $C_\perp(L/2)$, and $\chi_{s1}(L)$ as functions of $L$ are plotted in Fig. \ref{fig:surface}. The curves 
share similar slopes of the corresponding Y-c2 curves. 
Fitting these results according to Eqs. (\ref{cs1}), (\ref{cs2}), and (\ref{chis}), we obtain 
$y_{h1}=0.82(1)$, $\eta_{\parallel}=1.36(5)$ and $\eta_{\perp}=0.69(2)$, as listed in Tab. \ref{tab:scb}.
Again, they satisfy the scaling relations Eqs. (\ref{scalings1}) and (\ref{scalings2}), and
agree well with the exponents of the ordinary class in the 3D O(3) universality class.

\section{Discussion and Conclusion}
\label{sec:conclsn}

We have studied the spin-1/2 Heisenberg model on the 2D CDL lattice using quantum Monte Carlo simulations. 
We showed that the model realizes a 2D SPT Haldane phase when the ladders are weakly coupled. 
By tuning the interladder coupling, the model enters the N\'eel ordered phase first, then the trivial product RS phase,
through two quantum critical points. Furthermore, we demonstrated that the two QCPs are in the 3D O(3) universality class, no matter 
the magnetically disordered phase is a topologically nontrivial SPT phase or the simple product RS phase.

We have also studied the properties of the two gapped phases, including the finite-size behaviors of two 
topologically distinct string order parameters in the Haldane phase.
Compared with the previously studied models, this model has more abundant surface configurations. 
We showed that the Y surface, formed by the ends of the ladders, is gapless, while the X surface, exposed along 
the ladders, is gapful, in the Haldane phase.
Conversely, in the gapped RS phase, the Y surface is gapped, and the X surface is gapless. 

The mechanisms of the two gapless modes are different.
One is due to the properties of a topological SPT state.
The equivalent spin-1 chain of the diagonal ladder with free boundary conditions has a topological term of 
two spin-1/2 located at the boundaries. When the ladders are coupled to form a 2D system, the spin-1/2 excitations 
form a gapless edge state, according to the Lieb-Schultz-Mattis theorem. This explains
the gapless Y surface in the Haldane phase.
The other is due to the surface is formed by dangling spins. At least for spin-1/2 models, this can be understood
by assuming that the dangling spins form a spin-1/2
Heisenberg chain, which is gapless due to the topological $\theta$-term, suppressing the topological defects. 
This applies to the X surface in the RS phase.

We paid particular attention to the SCBs at the two bulk critical points. 
We have shown that the SCBs are always in the ordinary class on the surfaces that are gapped in the gapped bulk phases.
More importantly, we have demonstrated that nonordinary SCBs 
are realized at both critical points but only on the gapless surfaces of gapped bulk states
exposed by simply cutting bonds without fine-tuning the surface coupling, which is required to reach a 
multicritical point away from the ordinary class in the classical models. 

Considering that the gapless surface states in the gapped bulk phase are intimately related to the nonordinary SCBs at 
quantum critical points, and the mechanisms that lead to such gapless surface states are quantum mechanical, 
our work strongly supports the quantum origin of the nonordinary surface critical behaviors found in various quantum models.

At last, we would like to point out that: the nonordinary SCBs have been found in many quantum models, including various 
dimerized Heisenberg models, with spin-1/2 and spin-1, the 2D coupled spin-1 Haldane chains, and are now also found 
at two different critical points of the 2D coupled diagonal ladders, on two different surfaces respectively. The 
surfaces showing such nonordinary SCBs are
exposed by simply cutting lattices without any tuning of surface couplings. It is hard to believe all these systems
are, by chance, close to the special transition of the critical 3D classical O(3) model. 
Further investigation is called for.

\begin{acknowledgments}
We thank Prof. H.Q. Lin and Dr. Wenjing Zhu for valuable discussions.
This work was supported by the National Natural Science Foundation of China under Grant No.~12175015 and No.~11734002. 
The authors acknowledge support extended by the Super Computing Center of Beijing Normal University.
\end{acknowledgments}

\bibliography{ref.bib}

\clearpage
\appendix

\section{A summary of various results}

For the reader's convenience, the surface critical exponents of the coupled diagonal ladders(CDL), as well as other models 
  at critical points in the 3D O(3) universality class are listed in Tab.\ref{exp3} for comparison.
  Some field theoretical results from different methods are also listed for comparison.

\begin{ruledtabular}
  \begin{table*}[!t]
  \caption{  %Surface critical exponents of the coupled diagonal ladders (CDL
  For the reader convenience, the surface critical exponents of the coupled diagonal ladders(CDL), as well as other models 
  at critical points in the 3D O(3) universality class are listed for comparison, with
  %The results are mainly referred to \cite{Zhu2021}. 
  CHC the QCP of the coupled Haldane chains,
	  CD-DAF the Dimer-AF QCP of the columnar dimerized Heisenberg model, SD-DAF the Dimer-AF QCP of the staggered dimerized Heisenberg models,
	  DS-DAF and DS-PAF the Dimer-AF QCP and the PVBC-AF QCP of the dimerized Heisenberg model on the DS
	  lattice,respectively, 
	 3D CH  the three-dimensional classical Heisenberg model. 
	 For the types of surface configurations, D denotes dangling and N nondangling. 
	 "class." means classical model.
	 For the SCB class,  "Ord." is the abbreviation of ordinary, "Nonord." means nonordinary, and "Sp." special.
	  The field theoretical results(FT) from different methods are also listed for comparison. 
  }
  \label{exp3}
  \begin{tabular}{c l l l c c c} 
SCB class               &  Model/methods         &surfaces        &     Spin S         &$\eta_\parallel$       &  $\eta_\perp$   & $y_{h1}$ \\
  \hline
Nonord.                     & CDL                    &X-c2   &     1/2              & -0.56(1)              &   -0.259(3)      &  1.780(2)   \\
Nonord.                     &                     &Y-c1   &     1/2              & -0.511(2)              &   -0.237(2)      &  1.756(3)   \\
Ord.                     &                    &X-c1   &     1/2              & 1.36(6)              &   0.69(3)      &  0.82(1)   \\
Ord.                     &                    &Y-c2   &     1/2              & 1.318(31)              &   0.682(9)      &  0.852(46)   \\
\hline
Nonord.\cite{Zhu2021}                     & CHC                    &x surface   &     1              & -0.57(2)              &   -0.27(2)      &  1.760(3)   \\
Ord.\cite{Zhu2021}                    &                        &y surface   &     1              & 1.38(2)               &    0.69(2)      &  0.79(2)   \\ 
  \hline
Nonord. \cite{Ding2018}     & CD-DAF                 &  D         &     1/2            &  -0.445(15)           &    -0.218(8)    &1.7339(12)\\
Nonord.  \cite{Weber2018}      &                        &  D         &     1/2            &  -0.50(6)             &    -0.27(1)     &1.740(4) \\
Nonord.   \cite{Weber2019}     &                        &  D         &     1              &  -0.539(6)            &    -0.25(1)     & 1.762(3) \\
Ord.  \cite{Weber2018}     &                        &  N         &     1/2            &  1.30(2)              &    0.69(4)      & 0.84(1)    \\
Ord.  \cite{Ding2018}   &                        &  N         &     1/2            &  1.387(4)             &     0.67(6)     & 0.840(17) \\   
Ord.  \cite{Weber2019}     &                        &  N         &     1              &  1.32(2)              &    0.70(2)      & 0.80(1)    \\
\hline
Ord.  \cite{Ding2018}   &SD-DAF                  &  N         &      1/2           & 1.340(21)             &    0.682(2)     & 0.830(11) \\
     \hline
Nonord.  \cite{Zhang2017}   & DS-DAF                 & D         &      1/2           &   -0.449(5)           &   -0.2090(15)   & 1.7276(14) \\
Nonord.  \cite{Weber2018}      &                        & D         &      1/2           &   -0.50(1)            &   -0.228(5)     & 1.728(2)    \\
Ord.  \cite{Weber2018}     &                        & N         &      1/2           &   1.29(6)             &    0.65(3)      & 0.832(8)    \\
\hline
Ord.  \cite{Zhang2017}  & DS-PAF                  & N         &      1/2           &   1.327(25)           &    0.680(8)     & 0.810(20)\\
Ord.  \cite{Weber2018}     &                         & N         &      1/2           &   1.33(4)             &    0.65(2)      & 0.82(2)    \\
Nonord.   \cite{Weber2018}     &                         & D         &      1/2           &  -0.517(4)            &   -0.252(5)     &  1.742(1)  \\
          \hline
Ord.  \cite{Deng2005}       &3D CH                    &           &      class.        &                       &                 & 0.813(2)\\

\hline
Ord.\cite{Diehl1980}          & FT, 4-d $\epsilon$-exp  &           & class.             & 1.307                 &  0.664          & 0.846     \\
Ord.\cite{Diehl1986}           & FT, d-2 $\epsilon$-exp  &           & class.             & 1.39(2)               &                 &               \\
Ord.\cite{Diehl1994,Diehl1998}      & FT, Massive field       &           & class.             &  1.338                & 0.685           & 0.831      \\    
Ord.\cite{Gliozzi2015}         & FT, Conformal bootstrap &           & class.             &                       &                 & 0.831      \\           
Sp.\cite{Diehl1981}           & FT, 4-d $\epsilon$-exp  &           & class.             &  -0.445               & -0.212          &  1.723    \\     
  \end{tabular}
  \end{table*}
\end{ruledtabular}

%\end{CJK*}
\end{document}